# Non-Intrusive Signature Extraction for Major Residential Loads

Ming Dong, *Student Member, IEEE*, Paulo C. M. Meira, *Student Member, IEEE*, Wilsun Xu, *Fellow, IEEE*, and C. Y. Chung, *Senior Member, IEEE*

*Abstract*—This paper presents a technique to extract load signatures non-intrusively by using the smart meter data. Load signature extraction is different from load activity identification. It is a new and important problem to solve for the applications of non-intrusive load monitoring (NILM). For a target appliance whose signatures are to be extracted, the proposed technique first selects the candidate events that are likely to be associated with the appliance by using generic signatures and an event filtration step. It then applies a clustering algorithm to identify the authentic events of this appliance. In the third step, the operation cycles of appliances are estimated using an association algorithm. Finally, the electric signatures are extracted from these operation cycles. The results can have various applications. One is to create signature databases for the NILM applications. Another is for load condition monitoring. Validation results based on the data collected from three actual houses and a laboratory experiment have shown that the proposed method is a promising solution to the problem of load signature collection.

## I. Introduction

THE public's awareness of energy conservation has increased rapidly in recent years. At the same time, the development of smart grid, especially the vast deployment of smart meters enables users to access their energy data on a much more detailed scale [1]. This has promoted considerable interests in non-intrusive load monitoring (NILM) research [2]. It is hoped that NILM techniques will be able to decompose the energy consumption data into individual appliance level and thereby facilitate the energy saving decisions of customers.

To perform load decomposition, all NILM techniques must rely on the unique signatures of individual appliances [2]–[9]. At present, such signatures are collected intrusively such as directly measuring the sample appliances. Reference [10] recommends a current-sensor based registration device for signature collection. References [11] and [12] proposed a magnetic-sensor based event detector for the same purpose.

Manuscript received July 19, 2012; revised November 26, 2012; accepted January 29, 2013. This work was supported in part by Natural Sciences and Engineering Research Council of Canada (NSERC) and in part by São Paulo Research Foundation (FAPESP), Brazil. Paper no. TSG-00452-2012.

M. Dong and W. Xu are with the Department of Electrical and Computer Engineering, University of Alberta, Edmonton, AB T6G 2V4, Canada (e-mail: mdong@ualbeta.ca; wxu@ualberta.ca).

Paulo C. M. Meira is with the Department of Electrical Engineering, University of Campina, Campinas, Brazil (e-mail: meira@ualberta.ca).

C. Y. Chung is with the Department of Electrical Engineering, Hong Kong Polytechnic University, Hong Kong (e-mail: eecychun@polyu.edu.hk).

Color versions of one or more of the figures in this paper are available online at http://ieeexplore.ieee.org.

Digital Object Identifier 10.1109/TSG.2013.2245926

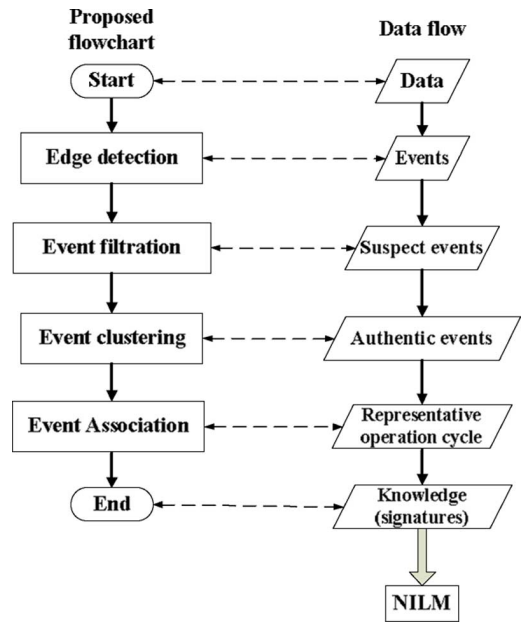

Fig. 1. Flowchart of proposed approach versus corresponding data flow.

It is clear that the above intrusive ways of signature collection are not desirable by ordinary house owners. Therefore, there is a need for methods that can collect the appliance signatures non-intrusively. If successful, unique signatures that are specific to an appliance in a particular home can be extracted and archived. Such signatures will make the identification of load activities more accurate in comparison to the cases where only generic signatures are available. The signature data can also be used for load condition monitoring purposes.

Non-intrusive identification of load activities and non-intrusive extraction of load signatures are two related but different problems. The signature extraction problem can be stated as follows: By using power consumption data for multiple days, establish the signatures that are specific and unique to the (major) appliances in the home under monitoring. To the best of our knowledge, non-intrusive extraction of load signatures has not been investigated by researchers as a distinct problem.

This paper presents our attempt to establish a non-intrusive load signature extraction method. When an appliance is turned on/off, a power change event is created and detected. The proposed method uses these events and derives knowledge (i.e., signatures) from them. As shown in Fig. 1, the method consists of five steps, event detection, event filtration, event clustering, event association, and signature extraction. The proposed method has been validated using various real life data and lab experiment.

## II. Event Filtration

The events caused by different appliances can be captured by using edge detection and this can be implemented based on the detection of discontinuities in the power level [13]. One operation cycle of each appliance may have two or more events depending on if it is an ON/OFF type appliance or a multi-state appliance. An ON/OFF type appliance such as a light bulb has only a pair of ON and OFF events while multi-state appliance such as furnace may have a series of operation state changes in the middle. An appliance's events are heavily characterized by its function and physical electric attributes and can be roughly located. For example, a fridge usually has an ON event with active power of 70–300 W and a reactive power of 30 to 200 Var. Due to the cooling function of a fridge, the ON events can be observed during 24 hours even when users are sleeping; A microwave usually has an ON event that has an active power of 800 to 2500 W and a heavy third harmonic content. As a cooking device, a microwave is likely to be observed before mealtime.

In this paper, the events that match the specific conditions of a certain appliance are defined as the "*suspect events*" of this appliance. The aim of event filtration is to locate the suspect events of a given appliance that may lead to the reconstruction of its entire operation cycle by carrying out future steps. To implement filtration, the conditions which can restrict the captured events are the: *active power range*, *reactive power range harmonic content range*, *with or without spike*, *single phase or double-phase* and *searching time*.

The active power, reactive power and harmonic content ranges are closely determined by the electric attributes of specific appliances. Residential loads can be roughly divided into four categories based on their linearity and reactivity: linear/active appliance, linear/reactive appliance, non-linear/active appliance and non-linear/reactive appliance. Depending on the type of category a certain appliance belongs to and its designed function, the numeric ranges of the above conditions can be quantified. For example, as a linear/active appliance, a resistive kettle has a very low reactive power and almost zero harmonic contents. Also, due to its water-heating purpose, its active power may statistically range from 1300 to 3000 W. Motor based appliances such as fridges can be viewed as inductors which lead to large reactive power. Another category is the switch-mode power supply based electronic appliances such as TVs and computers. They are neither inductive nor capacitive, but produce a large amount of harmonic contents. In addition, some appliances are both non-linear and reactive.

A Spike can be another ancillary condition that helps locate some induction-motor based appliance events. A large inrush current occurs at the first moment of operation when the rotor is triggered from the station into movement. This unique feature is accompanied by an ON event, which can be reflected as a sharp edge on the appliance's active power curve.

The phase condition can easily separate some events from other appliances' events. In North America, some appliances are connected between two hot phases to gain a 240 V voltage while others are connected between a single hot phase and a neutral to gain a 120 V voltage. From identification perspective, the events of double-phase appliances can be observed simultaneously at both hot phases, whereas the events of single-phase appliances occur at only one of them. In a residential house, only a few heavy-consumption appliances are connected to double-phase electric outlets. These appliances are the stoves, ovens and clothes dryers.

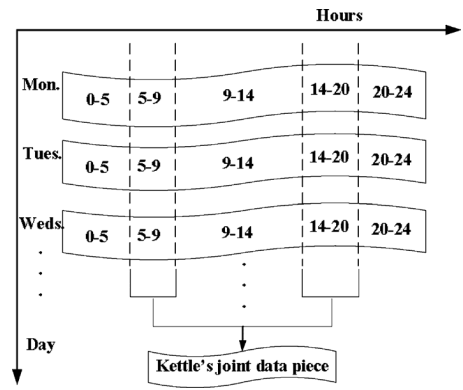

Fig. 2. Example of data piece connection for kettle.

The search window is another very important restriction that greatly reduces the searching space of suspect events. Some statistical studies are available on residential load behaviors which present typical appliance runtimes [15], [16]. For example, microwave's operations are more expected to be seen before breakfast, lunch and supper; lights are usually turned on in the early morning or after dark; fridges and furnaces are likely to run day and night. These occurrences show that, to locate the suspect events of specific appliances easily, it is best to search their less overlapped time ranges. For example, for a fridge, it is best to search from 2:00 AM–5:00 AM while many other appliances are generally inactive. For microwave, it is best to search the periods before meals. By doing so, the interference from other appliances can also be minimized.

In addition, to achieve the knowledge discovery purpose through massive amount of data, sufficient event samples must be obtained. Thus the data for multiple days need to be provided. According to the search window conditions, for different appliances, data pieces are cut from multiple days and re-joined together as a large data piece as shown in Fig. 2. It is expected that the objective appliances' events will have a much higher density in the joint data pieces.

Based on the above discussion, Table I presents an ON-Event filtration condition table for 10 major appliances. The listed P/Q/THD values can be understood as the generic ranges of these appliances in the geographic area of our research location (Edmonton, Canada). The signatures are based on the measurements of different brand/models of appliances from several residential houses. On average, 4 appliances of the same kind were measured. It should be noted that measuring all models/brands is impossible due to our limited resources. However, to compensate, the signature ranges are all expanded by at least 20% from our measurement values. For example, during the measurement, the power range of microwave was 1200–2000 W, and in Table I, it is modified to 800–2500 W in order to be more inclusive. Since the electric attributes of appliances are essentially determined by their functions, they will generally fall into the above ranges. However, in different countries or regions, different electricity voltage levels, climates and even cultures may affect the above signature ranges. Thus, a more local ON-Event

TABLE I
AN EXAMPLE OF ON-EVENT FILTRATION CONDITION TABLE

| Appliance | P(W) | Q(var) | THD (%) | With spike? | Phase Condition |
|---|---|---|---|---|---|
| Fridge | 70-300 | 30-200 | 0-20 | Yes | Single |
| Furnace | 120-800 | 200-800 | 0-20 | Yes | Single |
| Microwave | 800-2500 | 80-500 | 20-50 | No | Single |
| Stove (big element) | 1800-3000 | 0-30 | 0-5 | No | Double |
| Stove (small element) | 1000-2000 | 0-30 | 0-5 | No | Double |
| Oven | 2200-3600 | 0-30 | 0-5 | No | Double |
| Kettle | 1300-3000 | 0-30 | 0-5 | No | Single |
| Clothe dryer | 3000-6000 | 60-250 | 0-5 | Yes | Double |
| Washer (Front-load) | 80-300 | <100 | 65-95 | Yes | Single |
| Washer (Top-load) | 300-1000 | 300-1200 | 0-20 | Yes | Single |

filtration table can be updated according to the local measurements.

Appliances may have different working modes. For example, a stove usually has 4 heating elements (two big ones and two small ones) on its panel. The small elements of a stove consume only approximately half the power of the big elements. Since their signatures are quite different, they should be treated as two different types of appliances.

In Table I, P and Q are calculated based on the fundamental components of the current and voltage so that heavy harmonic contents will not affect their values [17]. For harmonic contents, the total harmonic distortion (THD) of current is used. Only the odd orders of the current harmonic contents ($i_k$) are considered due to their significance and usually a larger order than 9 does not need to be considered [17].

$$\text{THD} = \frac{\sqrt{\sum i_k^2}}{i_1}, k = 3, 5, 7 \ldots \quad (1)$$

According to the conditions listed in Table I, the suspect ON events of an appliance can then be identified by inspecting the events one by one in its joint data piece. To improve processing efficiency, only the ON events are verified and then considered as potential suspect events after passing inspection, because many transient features such as spikes accompany only ON events.

In order to guarantee that the events of an appliance can be included, the filtration conditions should not be set too strictly. In other words, the suspect events located by using the conditions in Table I may not be able to exclude the non-relevant events caused by other appliances; however, the conditions must allow enough of the events caused by the appliance to be included, because for a specific appliance, without enough event samples, the following procedures will fail or lead to inaccurate results.

## III. EVENT CLUSTERING

As discussed in Section II, the suspect events of appliance X are a group of ON-events that are possibly but not necessarily caused by X. Since the conditions are defined as loose ranges and the data for multiple days are used, other events which do not belong to X can be easily included in the suspect events occasionally. The aim of event clustering is to determine which suspect events are the real ones belonging to X. In this paper, the real events are named as the "*authentic events*" of X.

One basic assumption is that inside a suspect event group of appliance X, the number of events belonging to appliance X should be much larger than the number of events belonging to other appliances. The following examples below explain this assumption: suppose 50 suspect events of a fridge are located from a data piece for 2:00–5:00 AM of a whole week. It is possible that the user wakes up between 2:00–5:00 AM on Tuesday and for an unknown reason, uses an appliance such as a ceiling fan that accidentally meets the same filtration conditions as those of the fridge. Thus this fan's event will be mis-recognized as the fridge's suspect event and included in the fridge's 50-suspect-event group. This scenario might occur occasionally on certain days; however, this scenario is highly unlikely to occur as frequently as the fridge since such operations would be abnormal—he would have to turn on/off the fan not only every day but also as frequently as the fridge's kickins during sleeping time. Another example is a microwave. The user might turn on/off an unknown appliance that draws similar P/Q/Harmonic signals as a microwave does before mealtime, but it is unlikely to occur as often as a microwave. For any of these abnormal scenarios, the proposed unsupervised approach is not intended for and not able to deal with.

Furthermore, in some cases, corrupted events may occur that are usually due to simultaneous occurrences of more than one appliance event. This rarely happens but is possible. For example, the accidental overlapping of two single events belonging to a fridge and furnace will result in a power jump that is roughly the summation of the two events. Normally, the frequency of these corrupted events is much smaller than the authentic events.

Based on the above assumption, if the number of groups of events and the number of events each group possesses are known, the authentic events can be determined as those in the event group with the maximum number of members. Clustering is an effective tool to obtain the information. A clustering algorithm determines which events are roughly the same and can be grouped together as one cluster. After event filtration, since the clustering space is greatly reduced, many uncertainties and noises can be ruled out and clustering can be done much faster and more accurately compared to applying clustering to all the events of a day in the beginning. For instance, as Table II shows, suppose 88 suspect events of a fridge are located and after applying clustering algorithm, only four clusters are found. Then from the number of events each cluster owns, the dominant cluster which has 75 event members can be identified as having been caused by the fridge and its 75 event members can be labeled as the fridge's authentic events. On the other hand, smaller clusters are discarded because they are likely to be from other appliances or noises and cannot represent the objective appliance. The clusters and their mean values are listed in Table II. Two types of clustering methods are applicable to our study, as follows.

TABLE II
EXAMPLE OF COMPOSITION OF SUSPECT EVENTS

| Cluster Index | Mean P (W) | Mean Q (Var) | Mean THD(%) | Number of Events | Physical cause |
|---|---|---|---|---|---|
| 1 | 100.3 | 76.2 | 10.6% | 75 | Fridge |
| 2 | 87.7 | 67.9 | 10.3% | 10 | Fan |
| 3 | 73.6 | 58.8 | 2.2% | 2 | Motor X |
| 4 | 189.6 | 138.5 | 9.9% | 1 | Corrupted |

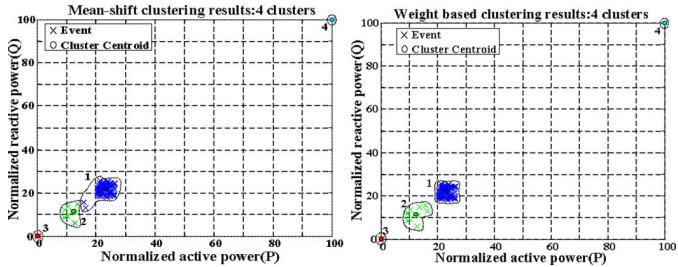

Fig. 3. Examples of two above clustering methods.

### A. Mean-Shift Clustering

Since the number of clusters (appliances) is not known in advance of clustering, many other clustering methods such as K-means are not applicable [18]. Mean-shift clustering [19], [20], however, is a robust non-parametric method that does not require prior information about the expected number of clusters. In our study, P/Q/THD are selected as three attributes used for clustering purpose. The "mean" linkage type is adopted to determine the distance between two clusters. All features are firstly normalized to the range of 1 to 100 through min-max scaling [21]. For mean-shift clustering, its bandwidth parameter is set from 5–20 due to the estimated variation of the electric signatures caused by the system voltage fluctuations.

### B. Weight-Based Clustering

Mean-shift clustering has one drawback due to the Euclidean space it uses where different attributes will be treated as equally important. However, as explained in Section II, depending on which category the appliance belongs to, different weights can be set for its three electric attributes. For example, for a fridge, which is a linear/reactive appliance, its active/reactive power features are more important than harmonic contents and should be emphasized; whereas for a stove, since its reactive power is negligible or could be caused by metering noises, either zero or a very low weight can be set for the reactive power feature. The proposed clustering algorithm has the following four steps:

**Step 1:** Initialize a random event as a one-element cluster;
**Step 2:** Compare other events with existing cluster mean(s) one by one. If an event is similar enough to a cluster mean, this event is merged into the same cluster and the cluster mean is updated. If an event is not similar to any of the existing cluster means, it is labeled as a new one-element cluster;
**Step 3:** Treat the existing cluster means as individual events and redo step 1 and 2. If two cluster means are similar enough, the two corresponding clusters are merged as a new cluster.
**Step 4:** Continue the iteration until no cluster is updated.

To quantify the similarity between an event and a cluster mean or two cluster means, the following formulas can be used: where $w_p, w_q, w_h$ in (2) are the weights assigned to the electric attributes $P, Q$ and THD according to appliance category information. Subscript $c$ indicates the cluster mean, and $e$ indicates a given event. (3), (4) and (5) can be used to calculate sub similarity indices $S_p$, $S_q$ and $S_h$. After overall similarity index $S$ is calculated, it can then be compared with a certain defined threshold such as 0.8 to determine if this event is similar enough to a cluster mean.

$$S = w_p S_p + w_q S_q + w_h S_h \tag{2}$$

$$S_p = \begin{cases} 1 - \frac{|P_e - P_c|}{P_c} & , |P_e - P_c| \leq P_c \\ 0 & \end{cases} \tag{3}$$

$$S_q = \begin{cases} 1 - \frac{|Q_e - Q_c|}{Q_c} & , |Q_e - Q_c| \leq Q_c \\ 0 & \end{cases} \tag{4}$$

$$S_h = \begin{cases} 1 - \frac{|\text{THD}_e - \text{THD}_c|}{\text{THD}_c} & , |\text{THD}_e - \text{THD}_c| \leq \text{THD}_c \\ 0 & \end{cases} \tag{5}$$

Generally, for the above two clustering methods, when the power variance inside the suspect event group is large, the bandwidth of mean-shift clustering can be lowered and the threshold of weight-based clustering can be increased in order to strengthen the cluster differentiation; when the power variance inside the suspect event group is small, the bandwidth of mean-shift clustering can be increased, and the threshold of weight-based clustering can be lowered in order to strengthen the cluster fusion.

To make a comparison, the mean-shift and weight-based clustering methods were applied to the 87 suspects discussed in Table II. They consist of 75 events from a fridge (1), 10 from a ceiling fan (2), 2 from a motor X (3) and 1 corrupted event (4). The clustering results are shown as P-Q 2D plots in Fig. 3. It shows that the mean-shift clustering method mistakenly includes 2 fans' events in its authentic group (the fridge's event group) because both the fridge and fan have similar THD values. However, because of the fridge's linear/reactive load type, its harmonic content should not contribute because it is usually small and instable, and can easily become mixed up with other linear appliances; in contrast, the weight-based clustering sets a very low weight for the harmonic content of a fridge and is able to exclude the 2 events caused by the fan. Generally, the proposed weight-based clustering has a better performance than mean-shift clustering in the proposed application.

### IV. EVENT ASSOCIATION

After the event clustering, the knowledge of the appliance is still incomplete since the other events except for the ON event in an appliance's operation cycle are still unknown. For accurate energy-tracking purposes, all the events and even the event pattern signatures [10] of an appliance have to be obtained. For example, for a fridge, its OFF event needs to be known; for multi-stage appliance such as a furnace, it is even more important since the ON event is only a small part of the furnace's full operation cycle. Its middle events and OFF event also need to be found. Knowing how these middle stage events occur in a particular pattern is also valuable. The overall purpose of event association is to determine the other events which also belong

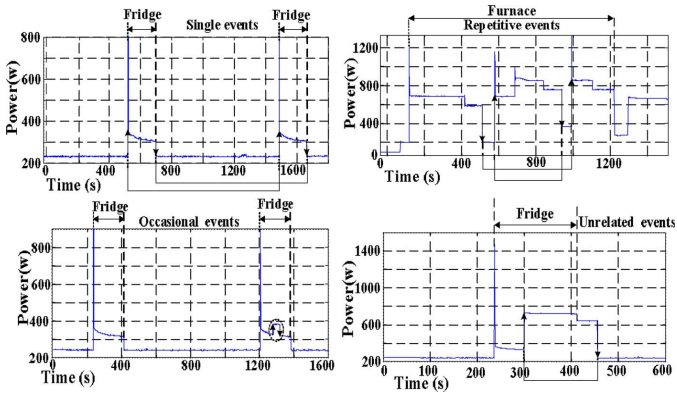

Fig. 4. Four types of associations of events.

TABLE III
AVERAGE DURATION AND DATA SEGMENT LENGTH FOR TYPICAL APPLIANCES

| Load name | Average duration | Data segment length |
|---|---|---|
| Fridge(cycle) | 15 mins | 22.5 mins |
| Furnace(cycle) | 20 mins | 30 mins |
| Microwave | 4 mins | 6 mins |
| Stove | 25 mins | 37.5 mins |
| Kettle | 4 mins | 6 mins |
| Oven | 10 mins | 15 mins |
| Washer | 45 mins | 67.5 mins |
| Clothes dryer | 50 mins | 75 mins |

TABLE IV
CRITERIA FOR ASSOCIATION DETERMINATION

| Association type | No. of events the cluster include (N) |
|---|---|
| Single event | cM<=N<=M |
| Repetitive event | N>=cM & n>=2 in at least one certain segment |
| Occasional event | bM<=N<cM |
| Unrelated event | N<bM (0<b<1) |

to the objective appliance and hence reconstruct its full *representative operation cycles*.

The proposed method counts the number of event occurrences and calculates the frequency to judge whether an event is associated with a specific appliance and determines the event's association type. An operation study of actual appliances [10] defines three association types:

- **Single event**: It has a strong association with an appliance. Once the appliance is working, this event must appear. However, it appears only once in single operation cycle. As Fig. 4 shows, a fridge has a fixed OFF event and it accompanies every operation cycle of the fridge but appears only once.
- **Repetitive event**: It has a strong association with an appliance. Once the appliance is working, this event must appear. It can appear more than once in a single operation cycle. As Fig. 4 shows, a furnace has repetitive events caused by its heating elements, which can be triggered multiple times according to the environment temperature.
- **Occasional event**: It has a weak association with an appliance. Once the appliance is working, this event might not occur. As shown in Fig. 4, when the door of a fridge is open, the light inside the chamber will be automatically switched on; when the door is closed, light will be switched off. Then users can see a small pair of power jumps between ON and OFF. An occasional event is usually caused by an ancillary element of appliance, such as the light of a fridge and the hood of a stove.
- **Unrelated event:** It has zero association with an appliance because it is caused by another appliance. As Fig. 4 shows, during a fridge's operation, a big incandescent lamp is also turned on. The ON event of the lamp happens before the OFF event of the fridge and the OFF event of the lamp happens after the OFF event of fridge.

The way to locate associated events and determine their association types can be determined by examining the data segments starting from the authentic ON events (determined by the event clustering in the last step). Each data segment is defined as a piece of data that has an authentic ON event in its beginning. The length of the piece is properly defined so it will not be too short (may fail to include all the associated events of the appliance) while it will also not be too long (may include too many unrelated events as interferences). The reasonable lengths can be defined as 1.5 times the average durations of the appliance operation. Table III shows the average durations of typical appliance operation and their segment lengths.

If there are M authentic ON events, there will also be M corresponding data segments, which may include not only the associated events of the appliance but also other unrelated events. The clusters of the events can be determined by applying the event-clustering methods discussed in Section III to the events in all segments together. To enhance the efficiency, the events with very different power levels than those of the authentic ON events are ruled out at the beginning before clustering. Because the former are not likely to be related. According to the definitions of three association types, theoretically, a single event should appear in each data segment and thus event number N should be equal to the data segment number M; for a repetitive event, since in some data segments, multiple events may appear, the event number N should be larger than M; for an occasional event, its occurrence should be more frequent than a threshold ruled by b. The factor b can differentiate whether or not an event is related to the objective appliance. Considering possible mistaken clustering or missing events due to improper data segmentation, the above theoretical criteria should be further corrected using confidence c and additional conditions as shown in Table IV. Normally, the values of b,c can be set as 0.3 and 0.8, respectively. An example on how event association works is shown in Section V.D.

## V. VERIFICATIONS AND DISCUSSIONS

### A. Verification and Discussions Based on Real House #1's Data

The above algorithms were tested by using data acquired from a real residential house in Edmonton, Canada for a week with no special attention from the house owner. A laptop-based data acquisition system was hooked to the house's electricity panel and continuously collected all the voltages and currents from the two hot phases (A and B) inside. It behaves just like a smart meter. The data were sampled every second. In each second, six consecutive cycles are acquired and each cycle has 256 points. Data acquired was then processed using the proposed algorithm and the operation cycles of specific appliances were automatically learned and reconstructed. In order to compare, operation times of appliance activities were manually recorded. By doing this, true operation cycles of appliances could be directly labeled as references. If the labeled events were more than one, their average values were used for comparison. In this house, there were 3027 events ($>50$ W) in Phase A, 2055 in Phase B and 2012 in Phase A-B. The total power of both phases on a typical day is shown in Fig. 5. Table I is adopted for event filtration. In total, 10 appliances were found and two different fridges were connected to the two different phases.

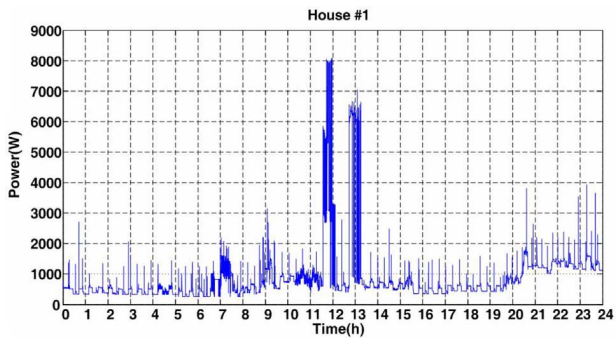

Fig. 5. The total power on a typical day from house #1.

TABLE V
SEARCH WINDOW AND RESULTS OF EVENT CLUSTERING

| Appliance | Search window | # of suspect events | # of clusters | Largest no. of events in single cluster |
|---|---|---|---|---|
| Fridge (PhaseA) | 2:00AM-5:00 AM | 69 | 10 | 21 (30.4%) |
| Fridge (Phase B) | 2:00AM-5:00 AM | 31 | 2 | 7(22.6%) |
| Furnace | 2:00AM-7:00AM 8:00PM-0:00AM | 4 | 1 | 4(100%) |
| Micro-Wave | 6:00AM-9:00AM; 11:00AM-2:00PM; 4:00PM-8:00PM | 54 | 7 | 27(50%) |
| Stove(big element) | 4:00PM-8:00PM | 517 (Repetitive) | 9 | 505(97.7%) |
| Stove (small element) | 4:00PM-8:00PM | 77 (Repetitive) | 5 | 60(77.9%) |
| Oven | 8:00AM-8:00PM | 7 | 5 | 3(42.9%) |
| Kettle | 6:00AM-9:00AM; 11:00AM-2:00PM; 4:00PM-8:00PM | 11 | 4 | 8(72.7%) |
| Clothes dryer | 8:00AM-11:00PM (Weekend) | 92 (Repetitive) | 5 | 61 (66.3%) |
| Washer | 3 hrs before dryer | 380 (Repetitive) | 10 | 189(49.7%) |

Table V shows the search window and authentic events determined in the step of event clustering. As can be seen, for each appliance, the number of authentic events (the largest no. of events in a single cluster) is larger than the average number of events per cluster. This result supports the assumption in Section III that authentic events will dominate inside suspect events.

As for the reconstructed operation cycles, the following two verification methods are conducted:

1) To visualize the effectiveness of the approach, the reconstructed operation cycles are compared with the labeled reference cycles in Fig. 6. For the sake of convenience, the events determined by using event clustering and association are aligned with the original events in the labeled cycles. As Fig. 6 shows, they are quite similar. The only difference is that some power variations between neighboring events are not captured, because the proposed approach focuses only on events. However, this missing information is not important since most NILM algorithms do not use it. Particularly, if any transient information accompanied with an event is needed, the corresponding event sample can be simply loaded for study.

2) An appliance's reconstructed operation cycle contains the electric signatures of all its associated events such as P, Q, THD (but not limited to these, depending on the needs of NILM). Table VI presents a detailed comparison between the signatures of individual events in the reconstructed cycles and labeled cycles. To compare, the mean values of event clusters are adopted. The errors are calculated with respect to the labeled reference events as true values. Table VI shows acceptable accuracy between the reconstructed cycles and labeled cycles. Please note the electric signatures of appliances can change within $\pm 5\%$ due to system voltage fluctuation. As Table VI shows, most of the active power errors are lower than 5%. Some errors of reactive power or harmonic THD are a little higher, because the corresponding appliances produce little reactive power or harmonic contents. The true values of these attributes are comparable to signal noises and can greatly fluctuate between measurements, especially when they are small. For example, even if several measurements are directly performed for the kettle's harmonic THD, differences at the listed level may still be observed. Another example is the washer. Since it is controlled by a variable speed driver, which continuously generates many repetitive event pairs with heavy noises (as shown in Fig. 6), the inherent signature consistency of these events is lower than that of other appliances. Thus, the above errors are quite normal. Furthermore, they will not affect NILM's identification significantly. Very small or zero weights will be given to these unstable attributes when making identification [10]. Similarly, classifiers such as neural networks [6] will "ignore" these signatures automatically during their training stages because these signatures in their training sets vary a lot too. The mean values being used can also balance out the variations of abnormal events inside clusters. Generally speaking, the electrical signature accuracy is very satisfactory.

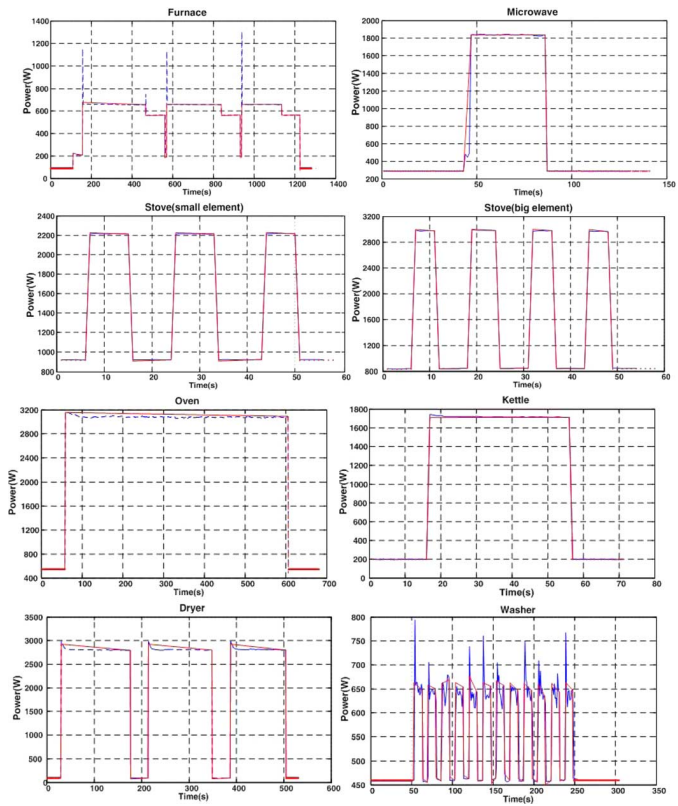

Fig. 6. Reconstructed cycles (red) vs. Labeled real cycles (blue) in house #1.

TABLE VI
ELECTRIC SIGNATURE ERROR BETWEEN RECONSTRUCTED CYCLES AND REFERENCE CYCLES FOR HOUSE #1

| Appliance Event | | Power Error % | Reactive Power Error % | Harmonic THD Error % |
|---|---|---|---|---|
| Fridge(PhaseA) | ON | 0.54 | 0.45 | -4.41 |
| Fridge(PhaseA) | OFF | 1.80 | -0.39 | -0.58 |
| Fridge(PhaseB) | ON | 3.00 | 7.31 | -4.75 |
| Fridge(PhaseB) | OFF | -0.01 | -0.04 | -5.48 |
| Furnace | ON | 0.73 | 4.11 | 1.94 |
| Furnace | Middle 1 | 1.13 | -4.91 | 2.16 |
| Furnace | Middle 2 | -1.41 | 1.40 | -3.16 |
| Furnace | Middle 3 | 1.11 | 1.44 | 6.57 |
| Furnace | OFF | -0.13 | 1.03 | -2.72 |
| Microwave | ON | -0.32 | 0.79 | -0.86 |
| Microwave | OFF | -0.85 | -2.72 | -2.68 |
| Stove(small) | ON | 0.52 | 0.17 | 2.24 |
| Stove(small) | OFF | -0.41 | 0.12 | -7.46 |
| Stove(big) | ON | -1.15 | 1.93 | -0.91 |
| Stove(big) | OFF | 0.14 | -4.70 | -0.68 |
| Oven | ON | 1.21 | -0.21 | 2.98 |
| Oven | OFF | -0.72 | -4.49 | -3.08 |
| Kettle | ON | -2.19 | 4.46 | -5.84 |
| Kettle | OFF | 0.24 | -2.81 | 12.08 |
| Dryer | ON | 0.96 | 0.97 | -0.20 |
| Dryer | OFF | 0.63 | -1.41 | -4.89 |
| Washer | ON | -4.21 | -0.54 | -1.41 |
| Washer | OFF | -5.31 | -3.32 | -7.29 |

### B. Verification and Discussions Based on Real House #2's Data

Similarly to house 1, another real residential house with a smaller family size in Edmonton was also measured for a week. The house contained 1451 events (>50 W) in Phase A, 1698 ones in Phase B and 564 ones in Phase A-B. The house's total power on a typical day is shown in Fig. 7. By using the proposed approach, 8 major appliances were found (the house has only one fridge). As Table VII shows, the electric signature is quite accurate with respect to the signatures contained in the labeled reference cycles.

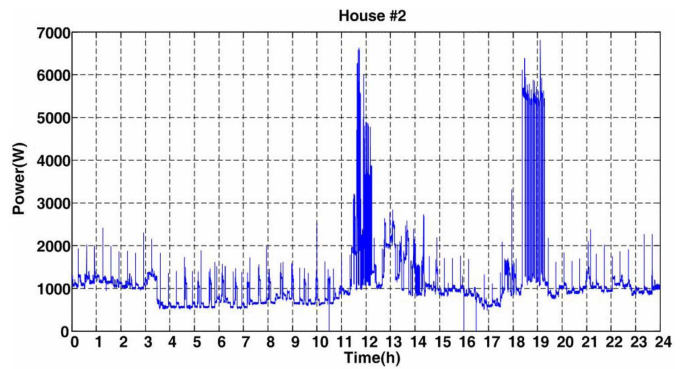

Fig. 7. The total power data on a typical day from house #2.

TABLE VII
ELECTRIC SIGNATURE ERROR BETWEEN RECONSTRUCTED CYCLES AND REFERENCE CYCLES FOR HOUSE #2

| Appliance Event | | Power Error % | Reactive Power Error % | Harmonic THD Error % |
|---|---|---|---|---|
| Fridge | ON | 1.84 | 3.04 | -5.87 |
| Fridge | ON | -1.93 | -1.01 | 3.33 |
| Furnace | ON | 0.82 | -2.03 | -5.57 |
| Furnace | Middle 1 | -1.71 | -6.99 | 1.20 |
| Furnace | Middle 2 | -4.37 | -5.54 | -5.39 |
| Furnace | Middle 3 | 1.42 | -0.40 | 3.93 |
| Furnace | Middle 4 | -0.80 | -0.86 | -2.55 |
| Furnace | OFF | -2.73 | -2.62 | -1.58 |
| Microwave | ON | -1.24 | -8.78 | -1.77 |
| Microwave | OFF | 1.91 | -6.46 | 3.25 |
| Stove(small) | ON | 4.65 | -5.41 | 2.02 |
| Stove(small) | OFF | 4.14 | -4.78 | 0.37 |
| Oven | ON | -1.42 | 2.88 | -5.58 |
| Oven | Middle | -1.47 | -4.60 | 2.85 |
| Oven | OFF | 1.04 | 0.44 | 1.69 |
| Kettle | ON | 1.04 | 0.44 | 1.69 |
| Kettle | OFF | 0.41 | 2.75 | -3.71 |
| Dryer | ON | 1.91 | -0.59 | -2.99 |
| Dryer | OFF | 0.97 | 0.98 | 3.33 |
| Washer | ON | 5.32 | 1.85 | 3.15 |
| Washer | OFF | 0.85 | 0.07 | 3.57 |

### C. Verification and Discussions Based on Real House #3's Data

To further verify the proposed approach, the data from house 3 in the REDD dataset [21] was used. REDD, a public dataset available for NILM research was released by MIT in 2011. The data was acquired from the greater Boston area in US. Please note: 1. In this dataset, many middle hours in a day are missing. For example, activities of the stove, oven and kettle were not recorded. 2. Although REDD labels specific appliances but unfortunately, some appliances are not clearly labeled such as the washer. In general, the dataset contains 1427284 seconds or roughly 16 days. The sampling rate was 275 points/cycle. The house contains 1160 events (>50 W) in Phase A, 2497 in Phase

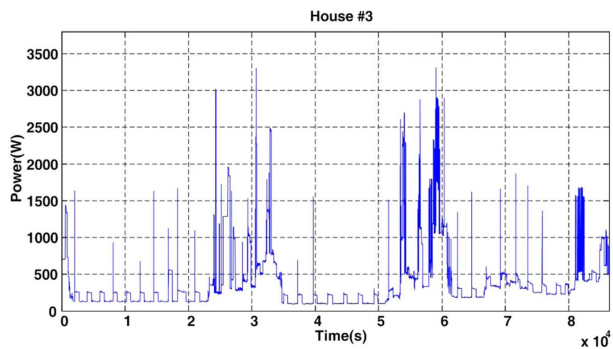

Fig. 8. The total power data of the first 86400 points from house #3.

TABLE VIII
ELECTRIC SIGNATURE ERROR BETWEEN RECONSTRUCTED CYCLES AND REFERENCE CYCLES FOR HOUSE #3

| Appliance | Event | Power Error % | Reactive Power Error % | Harmonic THD Error % |
|---|---|---|---|---|
| Fridge | ON | -1.04 | -1.23 | -4.16 |
| Fridge | ON | -0.20 | -0.55 | 0.25 |
| Furnace | ON | -2.36 | -2.49 | -1.92 |
| Furnace | Middle 1 | -1.47 | -5.46 | 5.39 |
| Furnace | Middle 2 | 7.21 | -1.81 | -0.96 |
| Furnace | Middle 3 | 2.46 | 3.88 | -7.12 |
| Furnace | OFF | 0.61 | 0.81 | -2.68 |
| Microwave | ON | -0.83 | 1.95 | -1.92 |
| Microwave | OFF | -0.56 | -3.20 | -3.36 |
| Dryer | ON | -1.96 | -4.32 | 3.25 |
| Dryer | OFF | -0.71 | 2.06 | -1.17 |
| Washer | ON | 1.08 | 4.39 | 1.94 |
| Washer | OFF | 1.83 | -1.69 | -8.03 |

TABLE IX
EVENT ASSOCIATION JUDGMENT FOR HEATER ($B = 0.3, c = 0.8$)

| Item | Number of apperance | Criteria used from Table IV | Event association type |
|---|---|---|---|
| Data segment | M=12 | --- | --- |
| Event 1 | N=12 | cM<=N<=M | Single event |
| Event 2 | N=21 | N>=cM & n>=2 | Repetitive event |
| Event 3 | N=21 | N>=cM & n>=2 | Repetitive event |
| Event 4 | N=12 | cM<=N<=M | Single event |
| Event 5 | N=12 | cM<=N<=M | Single event |
| Event 6 | N=4 | bM<=N<cM | Occasional event |
| Event 7 | N=4 | bM<=N<cM | Occasional event |
| Event 8 | N=3 | N<bM | Unrelated event |
| Event 9 | N=3 | N<bM | Unrelated event |
| Event 10 | N=2 | N<bM | Unrelated event |
| Event 11 | N=3 | N<bM | Unrelated event |

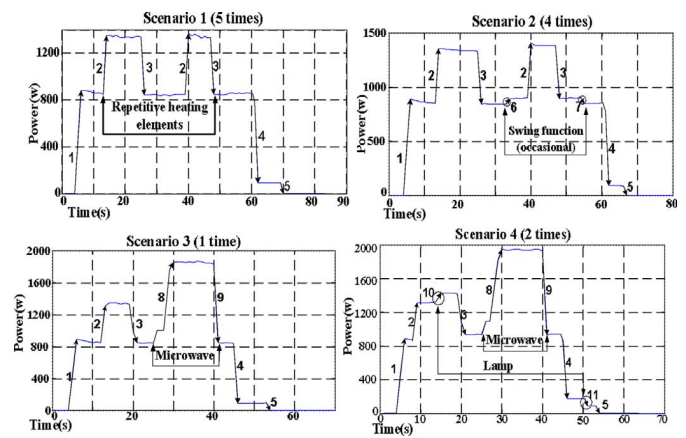

Fig. 9. Direct measurements of 4 scenarios of heater.

B and 523 in Phase A-B. As an example, the total power of the first 86400 seconds (the length of a day) is plotted in Fig. 8.

In spite of the above issues in the dataset, 5 major appliances can still be extracted using proposed approach. Surprisingly, even when the washer activities were not labeled in the REDD dataset, they were still able to be located and "mined" by using the proposed method. With the help of the proposed method, the washer's cycles were re-labeled.

As Table VIII reveals, the electric accuracy of reconstructed cycles is very good. It should be noted the filtration conditions in Table I used in this case are based on our local measurements in Edmonton, Canada. However, it is found still viable to deal with the house in the Boston area. This finding implies that appliances may share common signature ranges in similar geographic regions (such as northern part of North America).

### D. Verification of Event Association Based on Laboratory Data

To further test the effectiveness of the proposed event association approach specifically, a space heater was tested in the laboratory. To bring in the interference from other appliances' events, a lamp and a microwave were also connected to the same power supply bar which the space heater was also connected to. Then the aggregated signal of the power supply bar was measured under 4 types of scenarios, each for different times:

- **Scenario 1**: The heater ran with only its heating function on; this scenario was conducted 5 times.
- **Scenario 2**: The heater ran with both its heating function on and swaying function on (constantly changing the wind direction); this scenario was conducted 4 times.
- **Scenario 3**: The heater ran with only its heating function on; also, the microwave was switched on in the middle of the heater's operation; this scenario was conducted once.
- **Scenario 4**: The heater ran with only its heating function on; also, lamp and microwave were both switched on in the middle of the heater's operation; this scenario was conducted twice.

The above 4 scenarios simulate not only occasional events (such as those from the optional sway function) but also unrelated events caused by other appliances (in this case, microwave and lamp). In addition, the heating elements of the heater itself can be triggered repetitively. The purpose of this experiment was to test the capability of the algorithm to distinguish all types of event association for a given appliance. Examples of the measured power signals of the above 4 scenarios are plotted in Fig. 9. All events are also marked by different numbers according to their physical causes.

Event clustering was applied to the above four scenarios, which included 12 data segments in total. The clustering plots for both the ON and OFF events are shown in Fig. 10. The numbers marked around clusters are consistent with the numbers marked around events in Fig. 9. Also, according to the number of events each cluster owns, event association is judged as shown in Table IX. The table shows that the determined event pattern $1 \rightarrow 2 \rightarrow 3 \rightarrow 4 \rightarrow 5(6 \rightarrow 7)$ is completely consistent with the physical observations of the heater shown in Fig. 9 (∼∼∼: repetitive; bracket: occasional event).

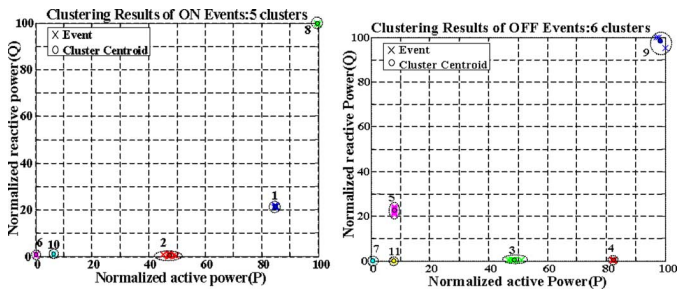

Fig. 10. Clustering results of all evens in space heater's 12 segments.

Overall, the above verification procedures and discussions show that the proposed signature extraction approach is capable of providing accurate electric and event pattern signatures for specific appliances automatically as long as the amount of feeding data is sufficient(more than a week). With this knowledge, most non-intrusive load identification algorithms [2]–[10] can be adopted without intrusive measurements for the training or registration stages. In other words, by combining proposed NISE with a NILM method such as [10], a truly non-intrusive load monitoring solution can be provided.

## VI. Conclusion

This paper addressed a novel problem related to the NILM research—the non-intrusive extraction of load signatures. Although most previous NILM studies addressed the identification or classification of load activities, the non-intrusive signature extraction of loads remains unsolved. An intrusive signature extraction process of loads can significantly impact the application of NILM to ordinary households. Thus, more research attentions should be drawn to this problem. The proposed approach is an unsupervised non-intrusive approach which can automatically extract load signatures by using the meter-side data and requires almost zero effort from users. The intention of this research was to eliminate or at least reduce the intrusive work load required by most existing NILM methods. The proposed approach uses event filtration, clustering and association to locate suspect events, determine authentic events and associated different events together to reconstruct operation cycle for an objective appliance, respectively. These reconstructed cycles contain most of the electric signatures and event pattern signatures of appliances and are enough for existing NILM approaches for identification and energy tracking purposes. The proposed approach was verified off-line by using the data acquired from 3 actual residential houses and a laboratory experiment. Both electric and event pattern signatures were tested, analyzed and discussed. Generally, the accuracy of the extracted signatures is satisfactory.

This paper focused on major appliances. Future research could study the smaller or unique appliances. Meanwhile, event filtration conditions can be updated with vast measurements in different geographic regions.

**Ming Dong** (S'08) received the B.Eng. degree in electrical engineering from Xi'an Jiaotong University, China, in 2008. He is currently pursuing the Ph.D. degree with electrical and computer engineering at the University of Alberta, Canada.

His research interests include smart grid power quality and computational intelligence applications in power system.

**Paulo C. M. Meira** (S'09) graduated in electrical engineering from University of Campinas (UNICAMP), Brazil, in 2007. He received the M.Sc. degree in electrical engineering from UNICAMP in 2010. He is currently pursuing the Ph.D. degree from the same university.

His research interests include smart grid applications, power systems reliability and protection.

**Wilsun Xu** (F'05) received the Ph.D. degree from the University of British Columbia, Vancouver, Canada, in 1989.

From 1989 to 1996, he was an Electrical Engineer with BC Hydro, Vancouver, Canada. Currently, he is with the University of Alberta, Edmonton, Canada, as a research chair professor. His current research interests are power quality, information extraction from power disturbances and power system measurements.

**C. Y. Chung** (M'01–SM'07) received the B.Eng. and Ph.D. degrees in electrical engineering from The Hong Kong Polytechnic University, Hong Kong, China, in 1995 and 1999, respectively.

After his Ph.D., he worked in the Electrical Engineering Department at the University of Alberta, Canada, and Powertech Labs, Inc., Surrey, Canada. Currently, he is the convener of the Power Systems Research Group and an Associate Professor in The Hong Kong Polytechnic University. His research interests include computational intelligence applications power system stability and planning.